\documentclass[11pt]{revtex4}
 
\usepackage{hyperref}
\usepackage{amsmath,amsfonts,amssymb}
\hypersetup{dvips,dvipdfm,colorlinks=true,urlcolor=magenta,filecolor=magenta,linktoc=page,citecolor=red,linkcolor=blue,bookmarks=true}
\usepackage{graphicx,epstopdf}

\begin{document}

\title{Homogeneous and Isotropic space-time, Modified torsion field and Complete cosmic scenario}

\author{Akash Bose$^1$\footnote {bose.akash13@gmail.com}}
% \author{Sourav Dutta$^2$\footnote {sduttaju@gmail.com}}
\author{Subenoy Chakraborty$^1$\footnote {schakraborty.math@gmail.com}}
\affiliation{$^1$Department of Mathematics, Jadavpur University, Kolkata-700032, West Bengal, India}
%$^2$ Department of Pure Mathematics, University of Calcutta, Kolkata-700019, west Bengal , India}

\begin{abstract}
	
The paper deals with cosmological solutions describing different phases of the Universe for the homogeneous and isotropic FLRW model of the Universe with torsion. Normally, torsion field is not suitable for maximally symmetric space time model. However, one may use a specific profile of vectorial torsion field, derived from a scalar function. By proper choices of the torsion scalar function, it is shown that a continuous cosmic evolution starting from the emergent scenario to the present late time acceleration is possible. Also thermodynamics of the system is analyzed and equivalence with Einstein gravity is discussed.

\end{abstract}

\maketitle


\section{Introduction}

Cartan in 1922 introduced an extension of Einstein's theory of gravity (known as Einstein-Cartan theory (ECT)), interpreting intrinsic angular momentum (i.e, spin) of the matter in terms of torsion \cite{cartan}-\cite{Cartan:1923zea}. Later in 1960s introduction of spin of the mater to general relativity \cite{Kibble:1961ba}-\cite{Sciama:1964wt}, showed ECT as the simplest classical modification of Einstein's theory [see \cite{Blagojevic:2013xpa} for a general review].

In ECT, the space-time contains asymmetric affine connection in contrast to general relativity having symmetric Christoffel symbols in Riemannian geometry. In particular, torsion is characterized the antisymmetric part of the affine (non-Riemannian) connection. As a result, the gravitational pull is not only described by the metric tensor but also by the independent torsion field. Geometrically, curvature has the tendency of bending the space-time while torsion twists it. In other words, parallel transport of a vector along a closed loop depends on the path due to the curvature of the space-time while presence of torsion may even prevent the formulation of the loop. Further, from physical point of view, presence of matter causes the curvature of the space-time while intrinsic angular momentum of matter characterizes the torsion.

%The maximally symmetric, homogeneous and isotropic FLRW (Friedmann-Lemaitre-Robertson-Walker) space-time model is the common choice for standard cosmology. In this model torsion field can be characterized by a specific profile \cite{Tsamparlis:1981xm} known as vectorial torsion field \cite{Olmo:2011uz}-\cite{Beltran}. In fact, space-time torsion and the associated matter spin are expressed by a time dependent scalar field. As a result, the symmetry of the Friedmann equations preserves the symmetry of the Ricci curvature tensor and the energy-momentum tensor.
On the other hand, the introduction of torsion field does not allow the space-time to be maximally symmetric (i.e. homogeneous and isotropic) which is the common choice for standard cosmology, rather it introduces anisotropic degrees of freedom. To overcome this unusual situation, a typical vectorial form of the torsion field has been introduced in recent past \cite{Tsamparlis:1981xm}. In this specific profile, the space-time torsion fields \cite{Olmo:2011uz}-\cite{Beltran} and the related matter spin are completely determined by a homogeneous scalar function. Further, this modified torsion field preserves the symmetry of the associated Ricci curvature tensor (in FLRW model) and hence the symmetry of Einstein tensor and energy-momentum tensor. Phenomenologically, torsion play the role of spatial curvature and it has an input in the terms of cosmological constant/ dark energy. So it is speculated that torsion may be responsible for accelerated expansion. However, it is observationally speculated that primordial nucleosynthesis is influenced by torsion and hence it is possible to constraint torsion field from observational point of view.

From an alternative view point, the effect of torsion in space-time can be interpreted as the intrinsic angular momentum of fermionic (i.e. spin) particles \cite{Poplawski:2009su}. Geometrically, it is related to the asymmetric affine connection of the space-time manifold. Hence matter field acts as a source for torsion and thereby enriching the cosmic descriptions. The well known Einstein-Cartan-Kibble-Sciama (ECKS) gravitational theory is very useful from the perspective of the invariance of local gauge in relation to the group of Poincare \cite{Hehl:1976kj}-\cite{Shapiro:2001rz}. It is to be noted that there is no observational evidence in favour of the existence of torsion field, rather there are suggestions for some experimental tests. 

Moreover, the standard cosmology in FLRW model has been a debating issue due to a series of observational evidences \cite{Goldstein:2002gf}-\cite{Eisenstein:2005su} for the last two decades. Attempts to resolve this issue has been continuing in two directions -- introduction of exotic matter (known as dark energy) in the frame work of Einstein gravity or modification of Einstein gravity theory itself. The second approach modifies the geometric part of the Einstein field equations and is termed as modified gravity theories. Essentially, a general form of the Lagrangian density than the usual Einstein-Hilbert action or some higher dimensional theories has been considered in this approach. The present torsion theory is an example of this type of modified gravity theory.

The present work shows an extensive study of cosmic evolution in FLRW model with torsion and it is examined whether torsion can be considered as an alternative to dark energy. A complete cosmic description has been presented for a continuous value of the torsion scalar function and thermodynamics of the system has been studied. The paper is organized as follows: section II deals with torsion in FLRW model, cosmic solutions and choice of torsion scalar function has been studied in section III. Section IV has discussed equivalence with Einstein gravity and non equilibrium thermodynamical prescription. The paper ends with a brief discussion in section V.
\section{Torsion and FLRW model}
The Einstein-Cartan theory which is based on the asymmetric affine connection of the space-time, instead of the Riemannian space, is the gravity theory with torsion. In particular, torsion tensor is described by the antisymmetric part of the affine connection, termed by 
\begin{equation}\label{eq1}
S^a_{bc}=\Gamma^a_{[bc]},
\end{equation}

which vanishes in absence of torsion. Since the metric tensor is covariantly constant ( i.e, $\nabla_c g_{ab}=0$), a generalized connection can be decomposed into a symmetric and an antisymmetric part as
\begin{equation}\label{eq2}
\Gamma^a_{bc}=\tilde\Gamma^a_{bc}+K^a_{bc},
\end{equation} 

where the symetric part $\tilde\Gamma^a_{bc}$ is the usual Christoffel symbols and the antisymmetric part $K^a_{bc}$ is termed as the contortion tensor with $K_{abc}=K_{[ab]c}$. This contortion tensor is related to the torsion tensor as
\begin{equation}\label{eq3}
K_{abc}=S_{abc}+2S_{(bc)a}.
\end{equation}

Thus one can think torsion as a connecting tool between the intrinsic angular momentum (spin) of the matter and the geometry of the space-time.

Due to antisymmetric nature of the torsion tensor one can define torsion vector as
\begin{equation}
S_a=S_{~ab}^b(=-S_{~ba}^b),
\end{equation}

and consequently, for the contortion tensor we have
\begin{equation}
K^b_{~ab}=2S_a=-K^{~~b}_{ab}
\end{equation}
%In non-zero space-time, matter and curvature are associated together by Einstein-Cartan field equation, namely
%\begin{equation}\label{eq3}
%R_{ab}-\frac{1}{2}Rg_{ab}=\kappa T_{ab}-\Lambda g_{ab}
%\end{equation}
%where $R_{ab}$ is the Ricci tensor, $R=R^a_a$ is the associated Ricci scalar and $T_{ab}$ is the energy momentum tensor. Here it is noted that both $R_{ab}$ and $T_{ab}$ are generally asymmetric due to presence of torsion.\\
%Now we consider the time like 4-velocity field $u_a$ (so $u_a u^a=-1$) which decompose the space-time into time and 3-dimensional space as $g_{ab}=h_{ab}-u_a u_b$, where $h_{ab}$ is the symmetric 3-tensor orthogonal to $u_a$ \cite{Tsagas:2007yx}.\\

In this work we consider homogeneous and isotropic FLRW space-time having line element
\begin{equation}\label{eq4}
ds^2 = -dt^2 +a^2(t) \left[\frac{dr^2}{1-Kr^2}+r^2 (d\theta^2+\sin^2\theta d\varphi^2)\right],
\end{equation}

where $a(t)$ is the scale factor with $H=\frac{\dot a}{a}$, H is the Hubble parameter and ` $\cdot{}$ ' represents differentiation with respect to cosmic time $t$ and $K$ is the curvature index. Also it is assumed that the Universe is consisting of perfect fluid with barotropic equation of state given by $p= \omega \rho$ and $\omega=\gamma-1$.

%The homogeneous and isotropic FLRW space-time model can not naturally incorporate a general form of the torsion field. Preserving the maximally symmetric nature of the 3D space-like hypersurface of the FLRW geometry the choice of the torsion field \cite{Tsamparlis:1981xm} has the following antisymmetric form
For spatially homogeneous and isotropic FLRW space-time torsion vector fully characterize the torsion tensor (and hence the contortion tensor). For this space-time the torsion vector can be written as \cite{Kranas:2018jdc}-\cite{Marques:2019ifg}
\begin{equation}
S_a=-3 \phi u_a,
\end{equation}

where $\phi=\phi(t)$ is a scalar function, $h_{ab}$ is the metric of the three space and $u_a$ is the four velocity field along the tangent to a congruence of time like curves. Then the torsion tensor and the contortion tensor can be expressed as
\begin{eqnarray}\label{eq5}
S_{abc}=2\phi h_{a[b}u_{c]},\\
K_{abc}=4\phi u_{[a}h_{b]c},
\end{eqnarray}

Clearly the torsion vector is a time like vector \cite{Pasmatsiou:2016bfv} and (the sign of) the scalar function $\phi$ indicates the relative orientation between the torsion and the 4-velocity (i.e, torsion vector is future directed for $\phi<0$ , while it is past directed for $\phi>0$). 

%So the Lagrangian can be written as
%\begin{equation}
%L(a,\dot{a},\phi,\dot{\phi})=6 a^3\left[\left(\frac{\dot{a}}{a}\right)^2+ \frac{K}{a^2}+2 \dot{\phi}+6\phi \left(\frac{\dot{a}}{a}\right)+4 \phi^2\right].
%\end{equation}

Further, the matter conservation equation in FLRW model takes the form \cite{Kranas:2018jdc}
\begin{equation}\label{eq10}
T_{a;b}^b=-4\phi T_{ab}u^b,
\end{equation}

which for perfect fluid has the explicit form
\begin{equation}\label{eq11}
\dot{\rho}+3(H+2\phi)(p+\rho)=4\phi\rho,
\end{equation}

where as usual $\rho$ is the energy density and $p$ is the thermodynamic pressure of the perfect fluid. Hence, the modified Friedmann equations due to torsion can be written as
\begin{eqnarray}
3H^2&=&\kappa \rho-3\frac{K}{a^2} -12\phi^2-12H\phi, \label{eq12}\\
2\dot{H}&=& -\kappa (p+\rho)+2\frac{K}{a^2} -4\dot{\phi}+8\phi^2+4H\phi. \label{eq13}
\end{eqnarray}
%and the continuity equation explicitly can be written as
%\begin{equation}
%\frac{\dot{\rho}}{\rho} = -3\gamma \frac{\dot{a}}{a}-2(1+3\omega)\phi,\label{eq14}
%\end{equation}

\section{Cosmic solutions and choices of torsion scalar function}

In this section, it is found that with proper choices of torsion scalar function, several cosmological solutions are possible in the present modified gravity theories. Throughout the work, the torsion scalar function is chosen in a typical (but general) form as
\begin{equation}\label{eq14}
\phi=-\lambda(a)H,
\end{equation}
%so $\dot{\phi}=-\lambda \dot{H}-\lambda ^\prime aH^2$,\\
%where ` $^\prime$ ' denotes differentiation with respect to scale factor $a$.\\

where $\lambda(a)$ is an arbitrary function of the scale factor. For simplicity, choosing $K=0$ i.e, flat space-time, the modified Friedmann equations (\ref{eq12}) and (\ref{eq13}) with the above choice (\ref{eq14}) for $\phi$ takes the form,
\begin{equation}
3H^2 (1-2\lambda)^2=\kappa \rho,\label{eq15}
\end{equation}
\vspace{-1cm}
\begin{equation}
2\dot{H} (1-2\lambda)=\kappa\rho \left[\frac{-3\gamma (1-2\lambda)^2+4(a\lambda^\prime+2\lambda^2-\lambda)}{3(1-2\lambda)^2}\right],\label{eq16}
\end{equation}

where ` $^\prime$ ' denotes differentiation with respect to scale factor $a$.

Now, combining equations (\ref{eq15}) and (\ref{eq16}) to eliminate $\rho$, one obtains the cosmic evolution equation as,
\begin{equation}
\frac{2\dot{H}}{3H^2}=\frac{4a\lambda^\prime}{3(1-2\lambda)}+2\lambda\left(\gamma-\frac{2}{3}\right)-\gamma. \label{eq17}
\end{equation}

\subsection{Emergent scenario: non singular universe}

In this section it will be examined whether it it is possible to have an emergent scenario (non-singular cosmological solution) for this theory. Now choosing $\lambda$ in such a way so that
\begin{equation}\label{eq18}
\frac{2a\lambda^\prime}{(1-2\lambda)}+3\lambda\left(\gamma- \frac{2}{3}\right) =\frac{\mu}{H},
\end{equation}

where $\mu$ is an arbitrary constant.

Equation (\ref{eq17}) takes the following form
\begin{equation}\label{eq19}
\dot{H}= \mu H-\frac{3\gamma}{2}H^2,
\end{equation}

Now depending on the signs of $\mu$ the possible solutions are as follows.

Case 1, $\mu > 0$,
\begin{eqnarray}\label{eq20}
\frac{H}{H_0}&=&\frac{\mu}{\frac{3\gamma}{2}H_0-\left(\frac{3\gamma}{2}H_0-\mu\right)e^{-\mu(t-t_0)}} ,\nonumber\\
\frac{a}{a_0}&=&\left[1+\frac{3\gamma H_0}{2\mu} \left(e^{\mu(t-t_0)}-1\right)\right]^{\frac{2}{3\gamma}}.
\end{eqnarray}

%and
%\begin{eqnarray}\label{eq21}
%\frac{H}{H_0}&=&\frac{\mu}{\frac{3\gamma}{2}H_0+\left(\mu-\frac{3\gamma}{2}H_0\right)e^{-\mu(t-t_0)}} ,\nonumber\\
%\frac{a}{a_0}&=&\left[1+\frac{3\gamma H_0}{2\mu} \left(e^{\mu(t-t_0)}-1\right)\right]^{\frac{2}{3\gamma}}.
%\end{eqnarray}
%
%according as $ \mu<\frac{3\gamma}{2}H_0$ and $\mu>\frac{3\gamma}{2}H_0$ respectively.

Case 2, $\mu = 0 $,
\begin{eqnarray}\label{eq22}
\frac{H_0}{H}&=&1+\frac{3\gamma}{2}H_0 (t-t_0), \nonumber\\
\frac{a}{a_0}&=&\left[1+\frac{3\gamma}{2}H_0 (t-t_0)\right]^{\frac{2}{3\gamma}}.
\end{eqnarray}

Case 3, $\mu < 0$ ,
\begin{eqnarray}\label{eq23}
\frac{H}{H_0}&=&\frac{\nu^2}{\left(\nu^2+\frac{3\gamma}{2}H_0\right)e^{\nu^2 (t-t_0)}-\frac{3\gamma}{2}H_0} ,\nonumber\\
\frac{a}{a_0}&=&\left[1+\frac{3\gamma H_0}{2\nu^2} \left(1-e^{-\nu^2(t-t_0)}\right)\right]^{\frac{2}{3\gamma}},
\end{eqnarray}

with $\mu=-\nu^2$.

In the above solutions $\lambda_0$, $t_0$, $a_0$ and $H_0$ are integration constants with $\lambda=\lambda_0$, $a=a_0$, $H=H_0$ at $t=t_0$.

Note that equation (\ref{eq20}) represents big bang singularity for $\mu<\frac{3\gamma}{2}H_0$, while for $\mu>\frac{3\gamma}{2}H_0$ the solution (\ref{eq20}) represents the emergent scenario of the universe and $\mu=\frac{3\gamma}{2}H_0$ represents only the inflationary era (i.e. the exponential expansion). The asymptotic limit for emergent scenario are the following:

$(i)~a\rightarrow a_0 \left(1-\frac{3\gamma H_0} {2\mu} \right)^{\frac{2}{3\gamma}},H\rightarrow0 ~\mbox{as}~t\rightarrow -\infty,$

$(ii)~a\sim a_0  \left(1-\frac{3\gamma H_0}{2\mu}\right) ^{\frac{2}{3\gamma}} , H\sim0~\mbox{as}~t\ll t_0,$
 
$(iii)~a\sim a_0 \left(\frac{3\gamma H_0}{2\mu}\right) ^ {\frac{2}{3\gamma}}e^{\frac{2\mu}{3\gamma} (t-t_0)} , H\sim\frac{2\mu}{3\gamma}~\mbox{as}~t\gg t_0. $

and the parameter $\lambda$ can be explicitly written $\left(\mbox{for~}\mu=3\gamma H_0\right)$ as
$$\frac{1}{1-2\lambda}=\frac{aH}{H_0} \left[\frac{1}{a_0(1-2\lambda_0)}+\left(\frac{3\gamma}{2}-1\right)H_0\int_{t_0}^{t} \frac{dt}{a(t)} \right]$$

Also (\ref{eq22}) and (\ref{eq23}) represent the big bang singularity and the big bang singularity occurs at the time $t=t_s$ given by,

$t_s=t_0+\frac{1}{\mu}\ln{\left| 1-\frac{2\mu}{3\gamma H_0}\right|} ~~\mbox{for the equation (\ref{eq20})},$

$t_s=t_0-\frac{2}{3\gamma H_0} ~~\mbox{for the equation (\ref{eq22})},$

$t_s=t_0-\frac{1}{\mu}\ln{\left|\frac{H_0}{H_0-\frac{2\mu}{3\gamma}}\right|} ~~\mbox{for the equation (\ref{eq23})}$.

\subsection{Different cosmological solutions and continuous cosmic evolution}

%In this section it is investigated whether this theory can be considered as a particle creation process in Einstein gravity and it is also examined whether a continuous cosmic evolution is possible for proper choices of parameter $\lambda$ for this particular choice of $\phi$. From the usual Friedmann equations with particle mechanism the cosmic evolution can be described as \cite{Chakraborty:2014oya}
%\begin{equation}\label{eq24}
%\frac{2\dot H}{3H^2}=\gamma\left (\frac{\Gamma}{3H}-1\right),
%\end{equation}
%where $\Gamma$ is the particle creation rate of the cosmic fluid particles. Comparing equation (\ref{eq24}) with equation (\ref{eq17}), one have
%\begin{equation}\label{eq25}
%\frac{\gamma\Gamma}{3H}= \frac{4a\lambda^\prime}{3(1-2\lambda)}+2\lambda \left(\gamma-\frac{2}{3}\right).
%\end{equation}
%The choices for the particle creation rate for these phases of evolution of the universe are :\\
%$(i)~\Gamma=3\mu_1 \frac{H^2}{H_1}$ (for inflationary era),\\
%$(ii)~\Gamma=3\mu_2 H$ (for matter dominated era),\\
%$(iii)~\Gamma=3\mu_3 \frac{H_2}{H}$ (for present accelerating phase).\\
%The deceleration parameter $q$ is given by
%$$q=-\left(1+\frac{\dot{H}}{H^2}\right).$$
%So the deceleration parameter $q$ for these three phases is given by
%\begin{equation}\label{eq26}
%q=\left\{
%\begin{array}{c c c}
%\frac{3\gamma}{2}-1-\mu_1\frac{H}{H_1}~~~\mbox{(for inflationary era)}~~~~~~~~~~~~~~ \\\\
%\frac{3\gamma}{2}-1-\mu_2~~~~~~\mbox{(for matter dominated era)}~~~~~\\\\
%\frac{3\gamma}{2}-1-\mu_3 \frac{H_2}{H^2}~~~\mbox{(for present %accelerating phase)}
%\end{array} \right.
%\end{equation}
Let $t_1$ be the time instant in which the universe evolves from inflationary era to matter dominated era. Similarly $t_2(>t_1)$ is the time instant at which the universe transits into late time acceleration era \cite{Chakraborty:2014oya}.\\   
\textbf{Inflationary era : }$(t<t_1)$ 

The choice for $\lambda$ is
$$\frac{1}{1-2\lambda}=1+\frac{H}{H_1}\left[ \frac{a}{a_1}\left(\frac{1}{1-2\lambda_1}-1+\frac{3\gamma \mu_1}{2}\right)-\frac{3\gamma \mu_1}{2}\right],$$

and the cosmic solution is given by
\begin{eqnarray}\label{eq24}
H&=&\frac{H_1}{\mu_1+(1-\mu_1)\left(\frac{a}{a_1}\right)^{\frac{3\gamma}{2}}},\nonumber\\
\mbox{i.e,}~~H&=&H_1\left[\mu_1\left\{LambertW\left(\frac{1-\mu_1}{\mu_1}\exp\left\{\frac{2(1-\mu_1)+3\gamma H_1 (t-t_1)} {2\mu_1}\right\} \right)+1\right\}\right]^{-1},\nonumber\\
a&=&a_1\left[\frac{\mu_1}{1-\mu_1}LambertW\left(\frac{1-\mu_1}{\mu_1}\exp\left\{\frac{2(1-\mu_1)+3\gamma H_1 (t-t_1)} {2\mu_1}\right\} \right)\right]^{\frac{2}{3\gamma}},\nonumber\\
q&=&\frac{3\gamma}{2}-1-\mu_1\frac{H}{H_1},
\end{eqnarray}

where $LambertW(x) e^{LambertW(x)}=x$ and $q=-\left(1+\frac{\dot{H}}{H^2}\right)$ is the deceleration parameter.\\
\textbf{Matter dominated era : }$(t_1<t<t_2)$

The choice for $\lambda$ is
$$\frac{1}{1-2\lambda}=\frac{3\gamma-2}{3\gamma(1-\mu_2)-2}+\frac{aH}{a_1 H_1}\left[\frac{1}{1-2\lambda_1}- \frac{3\gamma-2}{3\gamma(1-\mu_2)-2}\right],$$

and the cosmic solution can be written as
\begin{eqnarray}\label{eq25}
H&=&H_1\left(\frac{a}{a_1}\right)^{-\frac{3\gamma}{2}(1-\mu_2)}, \nonumber\\
\mbox{i.e,}~~H&=&H_1\left[1+\frac{3\gamma}{2}H_1 (1-\mu_2)(t-t_1)\right]^{-1},\nonumber\\
a&=&a_1\left[1+\frac{3\gamma}{2}H_1 (1-\mu_2)(t-t_1)\right]^{\frac{2}{3\gamma(1-\mu_2)}},\nonumber\\
q&=&\frac{3\gamma}{2}-1-\mu_2.
\end{eqnarray}
\textbf{Late time acceleration : }$(t>t_2)$

The choice for $\lambda$ is
$$\frac{1}{1-2\lambda}=\frac{aH}{H_2}\left[\frac{1}{ a_2(1-2\lambda_2)}-\left(\frac{3\gamma}{2}-1\right)H_2\int_{t_2}^{t} \frac{dt}{a(t)}\right],$$

and the cosmic solution is in the form of
\begin{eqnarray}\label{eq26}
H&=&H_2\left[\mu_2 +(1-\mu_2) \left(\frac{a}{a_2}\right)^{-3\gamma} \right]^{\frac{1}{2}},\nonumber\\~~\mbox{i.e,}~~
H&=&\sqrt{\mu_2}H_2\coth{\left\{\frac{3\gamma}{2}\sqrt{\mu_2}H_2(t-t_i)\right\}} \nonumber \\
a&=&a_2\left[\sqrt{\frac{1-\mu_2}{\mu_2}}\sinh\left\{\frac{3\gamma}{2}\sqrt{\mu_2} H_2 (t-t_i)\right\}\right]^{\frac{2}{3\gamma}},\nonumber\\
q&=&\frac{3\gamma}{2}-1-\mu_2 \left(\frac{H_2}{H}\right)^2.
\end{eqnarray}

where $t_i$ is the constant of integration.
Here $(\lambda_1,a_1,H_1)$ and $(\lambda_2,a_2,H_2)$ are the values of parameter $\lambda$, scale factor and Hubble parameter respectively at the transition points $t=t_1$ and $t=t_2$.

It will be now examined whether this cosmic evolution across these three phases is continuous or not. Then continuity of the deceleration parameter at the transition time $t=t_1$ and $t=t_2$, gives $$\mu_1 =\mu_2.$$

Continuity of physical parameters (i.e Hubble parameter, scale factor) at $t=t_1$ is obvious while the continuity at $t=t_2$ provides the following relations : 
\begin{eqnarray}\label{eq27}
\frac{3\gamma}{2}(1-\mu_2)(t_2-t_1)=\frac{1}{H_2}-\frac{1}{H_1}~~~~,~~~~\left(\frac{a_1}{a_2}\right)^{\frac{3\gamma(1-\mu_2)} {2}}=\frac{H_2}{H_1},
\end{eqnarray}
\begin{equation}\label{eq28}
\frac{1}{1-2\lambda_2}=\frac{3\gamma-2}{3\gamma(1-\mu_2)-2}+\frac{a_2 H_2}{a_1 H_1}\left[\frac{1}{1-2\lambda_1}- \frac{3\gamma-2}{3\gamma(1-\mu_2)-2}\right],
\end{equation}
\begin{equation}
\sinh(\eta_2)=\sqrt{\frac{\mu_2}{1-\mu_2}},
\end{equation}\label{eq29}
\mbox{where}
\begin{equation}
\eta_2=\frac{3\gamma}{2}\sqrt{\mu_2} H_2 (t_2-t_i).\nonumber
\end{equation}
\begin{figure}
	\begin{minipage}{0.42\textwidth}
		\centering\includegraphics[width=0.86\textwidth]{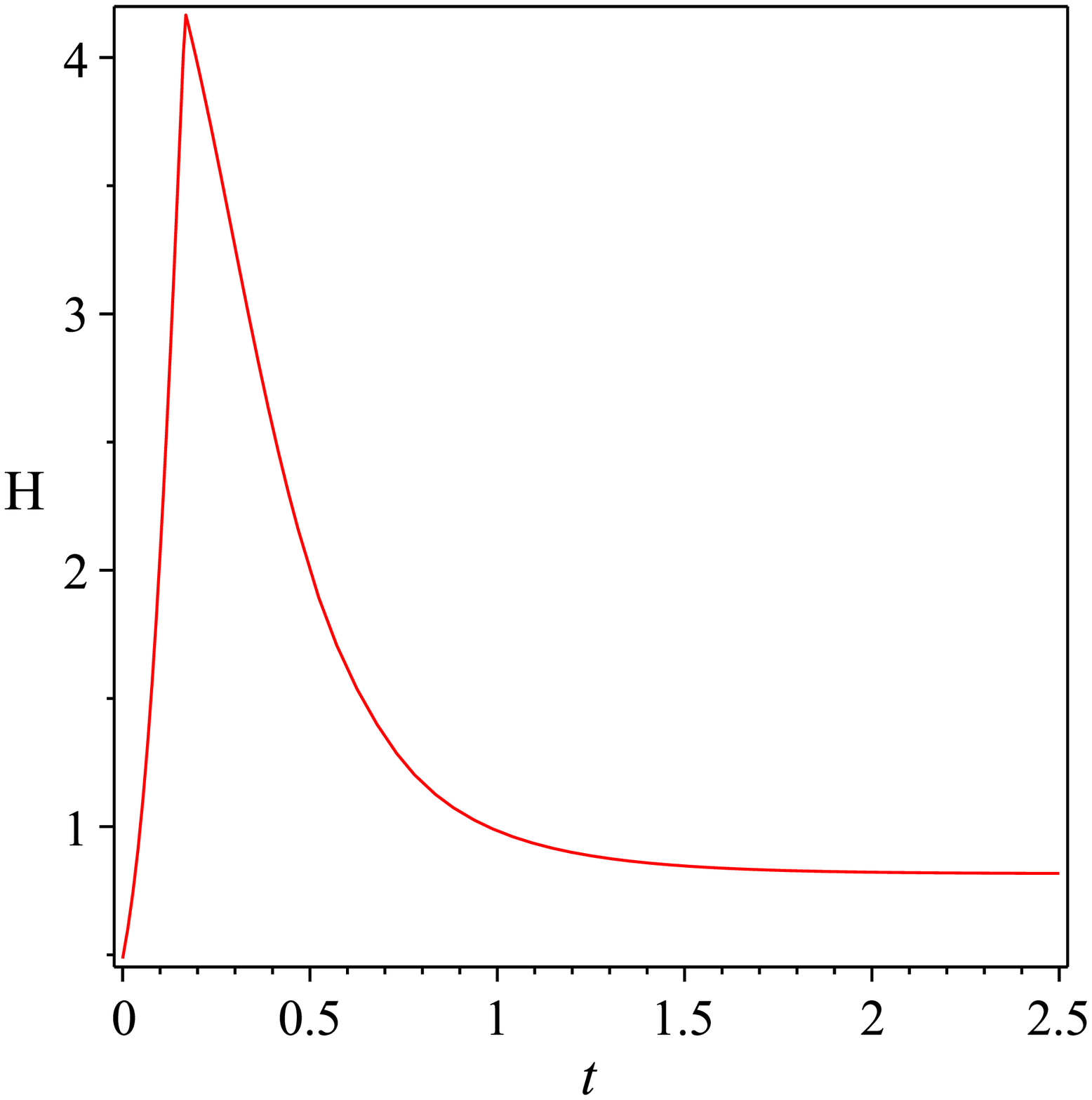}\\
		\caption{The Hubble parameter $H(t)$ is plotted against $t$}
		\label{fig1}
	\end{minipage}
	$\ \ \ \ \ \ \ \ \ \ \ \ \ \ \ \ \ \ \ \ \ $
	\begin{minipage}{0.42\textwidth}
		\centering\includegraphics[width=0.86\textwidth]{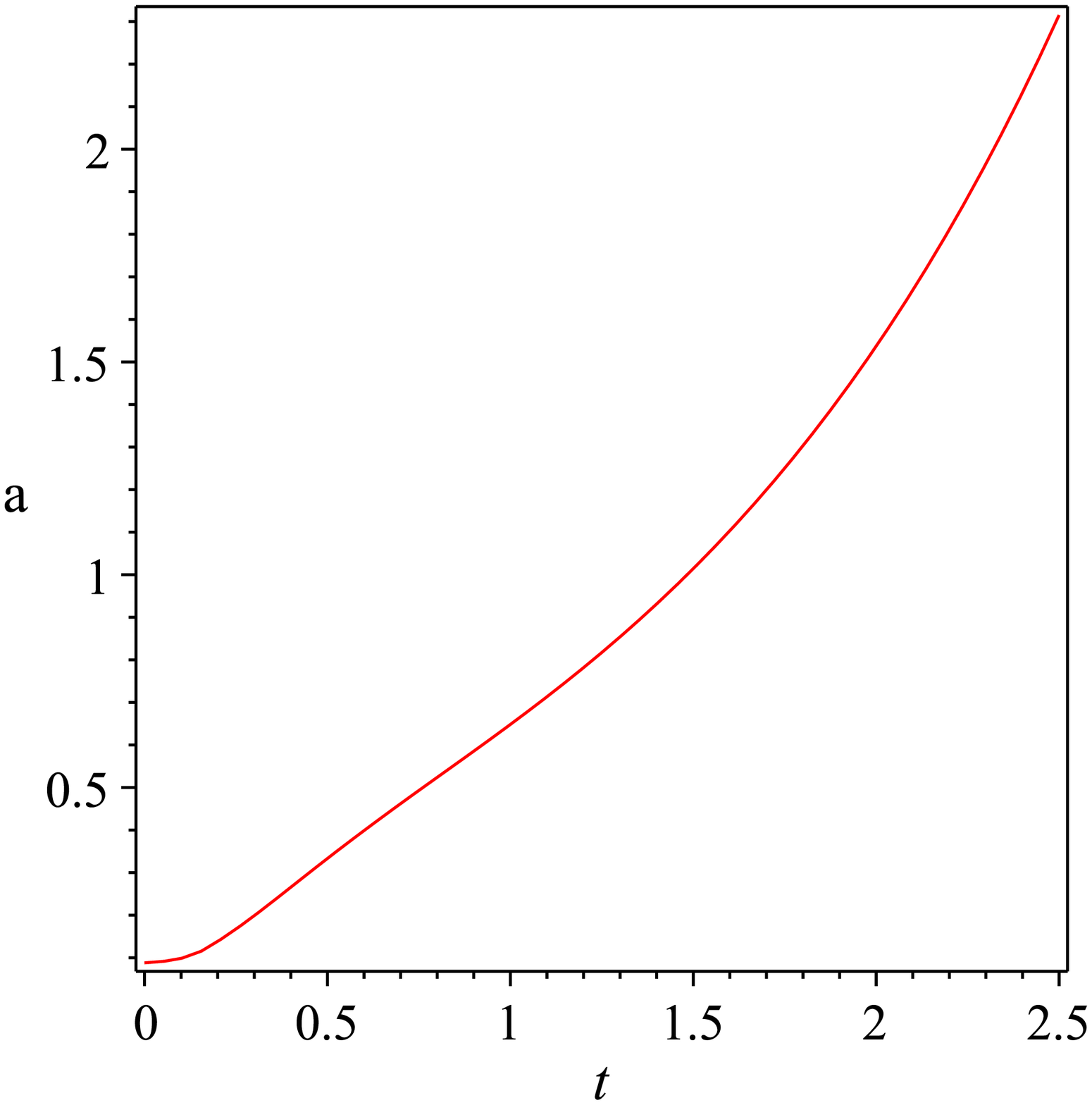}\\
		\caption{The evolution of the scale factor with respect to time $t$}
		\label{fig2}
	\end{minipage}
	%\end{figure}
	%\begin{figure}
	\begin{minipage}{0.42\textwidth}
		\centering\includegraphics[width=0.86\textwidth]{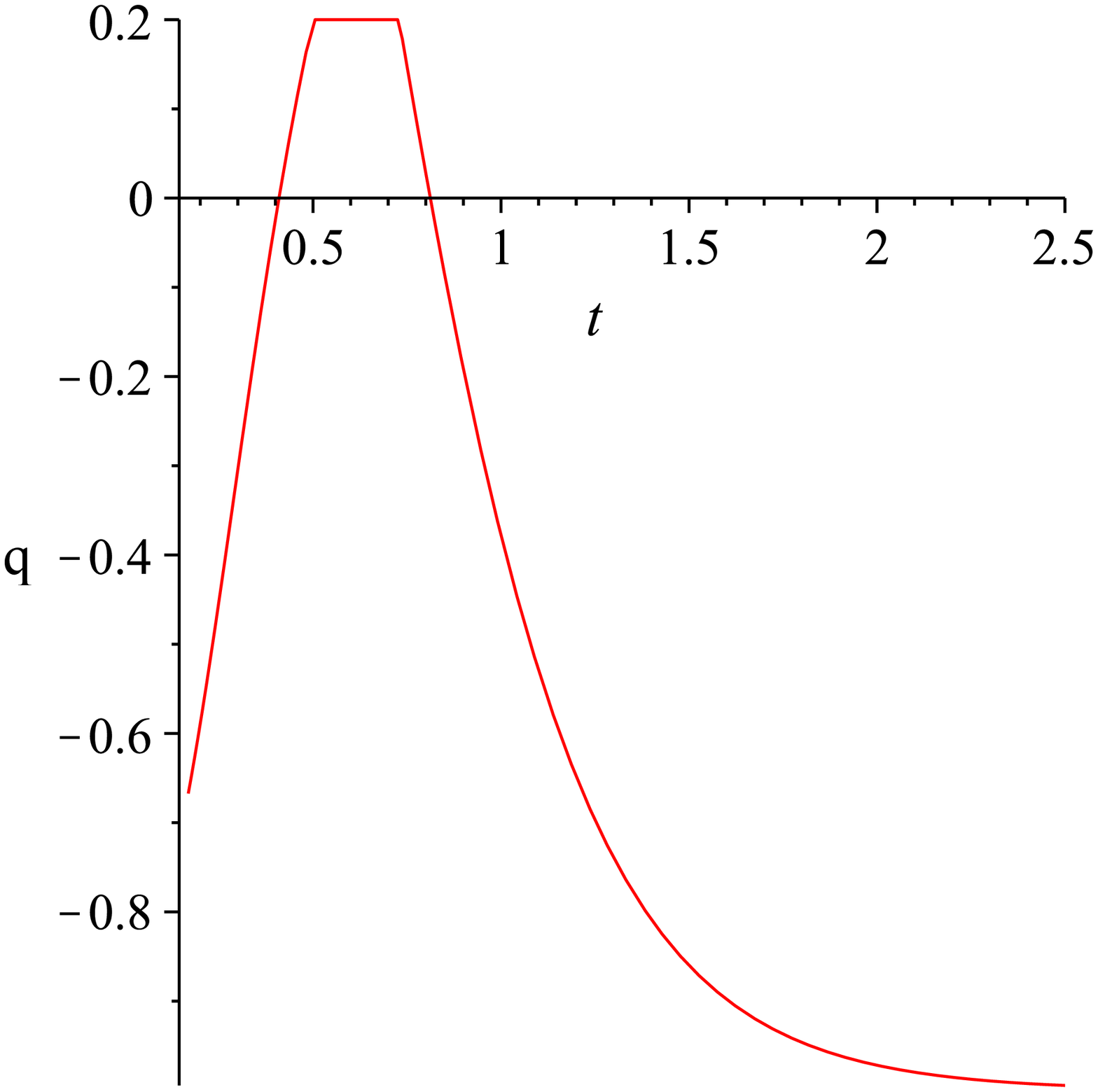}\\
		\caption{The deceleration parameter $q$ is plotted against $t$.}
		\label{fig3}
	\end{minipage}
	$\ \ \ \ \ \ \ \ \ \ \ \ \ \ \ \ \ \ \ \ \ $
	\begin{minipage}{0.42\textwidth}
		\centering\includegraphics[width=0.86\textwidth]{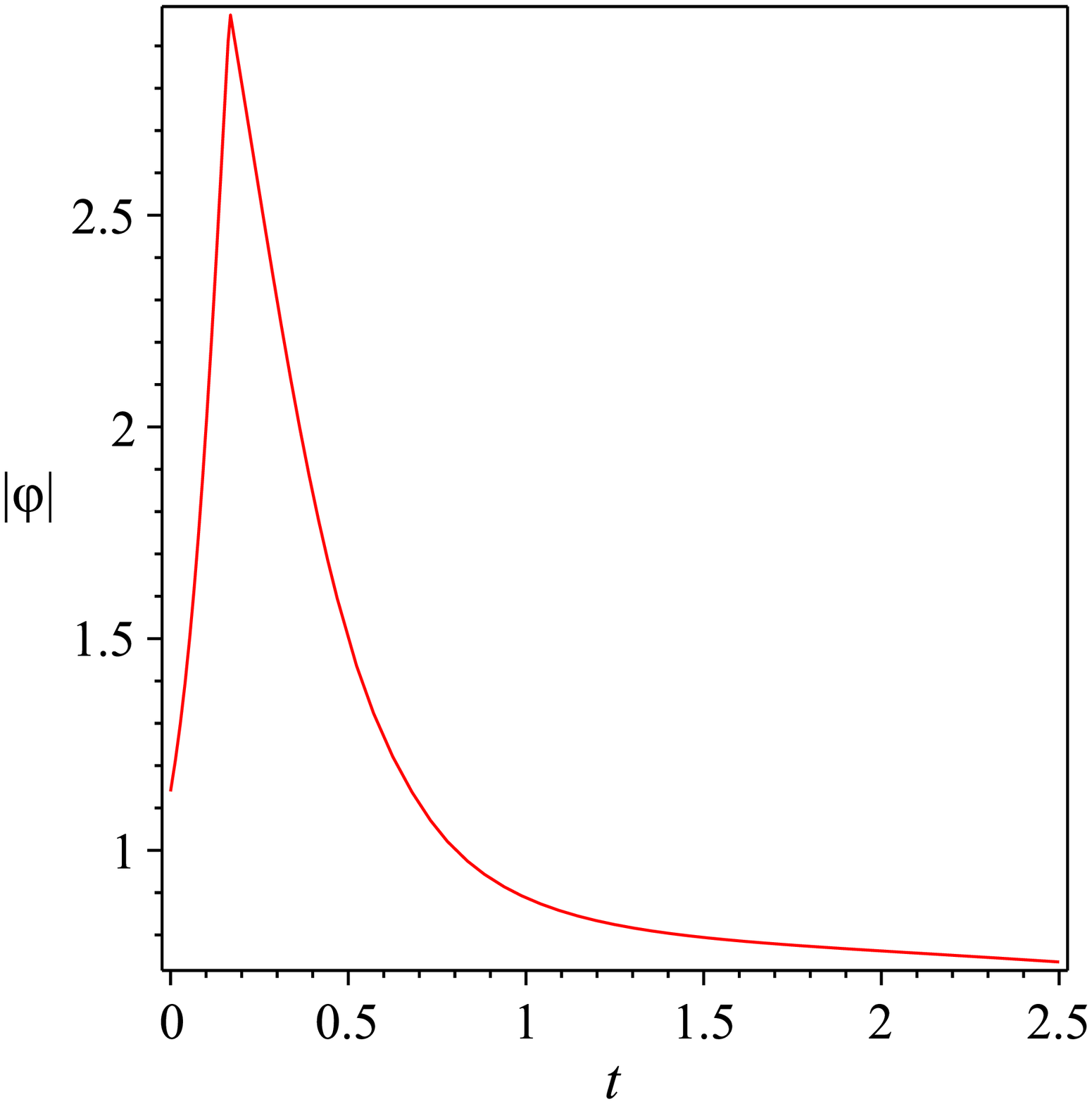}\\
		\caption{The evolution of the modulus of torsion $\phi$ with respect to time $t$.}
		\label{fig4}
	\end{minipage}
\end{figure}
Also let $t=t_0(<t_1)$ be the time instant in which the universe makes a transition from emergent scenario to inflation. Then from the continuity across the transition time $t=t_0$, we have
\begin{equation}\nonumber
\frac{1}{1-2\lambda_0}=1+\frac{H_0}{H_1}\left[ \frac{a_0}{a_1}\left(\frac{1}{1-2\lambda_1}-1+\frac{3\gamma \mu_1}{2}\right)-\frac{3\gamma \mu_1}{2}\right],
\end{equation}
\begin{equation}\label{eq30}
\frac{H_0}{H_1}=\mu_1+(1-\mu_1)\left(\frac{a_0}{a_1}\right)^{\frac{3\gamma}{2}},
\end{equation}

where $\lambda_0$, $H_0$ and $a_0$ are the values of parameter $\lambda$, Hubble parameter and scale factor at $t=t_0$ respectively. Also the smoothness of physical parameters $H$, $a$, $q$ ans $\phi$ are shown graphically in FIG. (\ref{fig1}) - (\ref{fig4}) for the parameter values $\mu_1 = 0.4$, $\gamma = \frac{4}{3}$, $H_1 = 2$, $a_1=\frac{1}{3}$, $H_2=1.29$, $t_1=0.5$, $t_0=\frac{1}{6}$, $\lambda_1=\frac{3}{4}$ and $\mu=3\gamma H_0$.

\section{Equivalence to Einstein gravity and non-equilibrium thermodynamical prescription}

In this section the thermodynamics of the gravity theory with torsion has been discussed in FLRW model. Also it is checked that, is this theory equivalent to the particle creation in Einstein gravity and temperature of the fluid particles is evaluated.
Equation (\ref{eq12}) and (\ref{eq13}) can be written as 
\begin{eqnarray}
3H^2&=&\kappa\rho-12\phi^2-12H\phi\nonumber\\&=&\kappa(\rho+\rho_e)=\kappa\rho_T,\label{eq31}\\
2\dot{H}&=&-\kappa\gamma\rho-4\dot{\phi}+8\phi^2+4H\phi\nonumber\\&=&-\kappa\left[(p+\rho)+(p_e+\rho_e)\right]=-\kappa (p_T+\rho_T),\label{eq32}
\end{eqnarray}

where $\rho_e,p_e$ are energy density and thermodynamic pressure of the effective fluid particles and are defined as,
\begin{eqnarray}
\kappa \rho_e&=&-12\phi^2-12H\phi,\label{eq33}\\
\kappa p_e&=&4\phi^2+8H\phi+4\dot{\phi}.\label{eq34}
\end{eqnarray}

From the field equations (\ref{eq31}) and (\ref{eq32}) due to Bianchi identity one has,
\begin{equation}
\dot{\rho_T}+3H(p_T+\rho_T)=0.\label{eq35}
\end{equation}

From equations (\ref{eq11}) and (\ref{eq35}) one have the individual matter conservation equations as,
\begin{eqnarray}
\dot{\rho}+3H(p+\rho)&=&Q,\label{eq36}\\
\dot{\rho_e}+3H(p_e+\rho_e)&=&-Q.\label{eq37}
\end{eqnarray}

Thus the modified Friedmann equations can be interpreted as Friedmann equations in Einstein gravity for an interacting two fluid system of which one is the usual normal fluid under consideration and other is the effective fluid and the interaction term is given by $Q=-2(\rho+3p)\phi.$

For interacting two fluid system the interacting term $Q$ should be positive as the energy is transferred to usual fluid. In this case $Q>0\implies\phi<0$.

One can write the above conservation equations (\ref{eq36}) and (\ref{eq37}) in terms of state parameter as 
\begin{eqnarray}
\omega^{(eff)}&=&\omega+2(1+3\omega)\frac{\phi}{3H},\label{eq38}\\
\omega^{(eff)}_ e &=&\omega_e-2(1+3\omega)\frac{\phi}{3H}r, \label{eq39}
\end{eqnarray}

where $r=\frac{\rho}{\rho_e}$ is coincidence parameter and $\omega_e=\frac{p_e}{\rho_e}$ is equation of state parameter of effective fluid.

Equations (\ref{eq36}) and (\ref{eq37}) can also be written as
\begin{eqnarray}
\dot{\rho}+3H(p+\rho+p_c)&=&0\label{eq40},\\
\dot{\rho_e}+3H(p_e+\rho_e+p_{ce})&=&0,\label{eq41}
\end{eqnarray}

with
\begin{eqnarray}
 p_c&=&\frac{2\phi}{3H}(\rho+3p)=-p_{ce},\label{eq42}
\end{eqnarray}

where $p_c$, $p_{ce}$ are the dissipative pressure of the fluid components.

In non-equilibrium thermodynamics this dissipative pressure may be caused by particle creation process. So the particle number conservation equations take the form,
\begin{eqnarray}
\dot{n}+3Hn&=&\Gamma n,\label{eq43}\\
\dot{n_e}+3Hn_e&=&\Gamma_e n_e.\label{eq44}
\end{eqnarray}

Here $n$ denotes the normal fluid particles density and $n_e$ represents the number density of effective fluid particles.

If we assume the non-equilibrium thermodynamical process to be adiabatic then the dissipative pressures are related to the particle creation rate linearly as \cite{Pan:2014lua},
\begin{eqnarray}\label{eq45}
p_c&=&-\frac{\Gamma}{3H}(p+\rho),\nonumber\\
p_{ce}&=&\frac{\Gamma_e}{3H}(p_e+\rho_e).
\end{eqnarray}

Comparing equations (\ref{eq42}) and (\ref{eq45}) we have,
\begin{eqnarray}
\Gamma&=&-2\phi \frac{\rho+3p}{\rho+p},\label{eq46}\\
\Gamma_e&=&2\phi \frac{\rho+3p}{\rho_e+p_e}.\label{eq47}
\end{eqnarray}

Thus the particle creation rate directly related to $\phi$. It is also clear that $\Gamma>0$ i.e, usual fluid particles are created and $\Gamma_e<0$ i.e, effective particles are annihilated. Also $p_{ct}=p_c+p_{ce}=0$ so the particle creation rate for resulting fluid particles vanishes identically. Hence resulting fluid forms a closed system.

Further, because of particle creation mechanism there is an energy transfer between the two fluid systems. So these two systems may have different temperatures.

Using Euler's thermodynamical equation, the evolution of the temperature of the individual fluid is given by \cite{Saha:2014uwa},
\begin{eqnarray}
\frac{\dot{T}}{T}&=&-3H\left(\omega^{(eff)}+\frac{\Gamma}{3H}\right)+\frac{\dot{\omega}}{1+\omega},\label{eq48}\\
\frac{\dot{T_e}}{T_e}&=&-3H\left(\omega^{(eff)}_e-\frac{\Gamma_e}{3H}\right)+\frac{\dot{\omega}_e}{1+\omega_e},\label{eq49}
\end{eqnarray}

where $\omega^{(eff)} $ and $\omega^{(eff)}_e $ defined in equations (\ref{eq38}) and (\ref{eq39}) can be written in terms of particle creation rate as,
\begin{eqnarray}
\omega^{(eff)}&=&\omega-\frac{\Gamma}{3H}(1+\omega),\label{eq50}\\
\omega^{(eff)}_e&=&\omega_e+\frac{\Gamma_e}{3H}(1+\omega).\label{eq51}
\end{eqnarray}

Integrating equations (\ref{eq48}) and (\ref{eq49}) we have,
\begin{eqnarray}
T&=&T_0 (1+\omega)exp\left[-3\int_{a_0}^{a} \omega \left(1-\frac{\Gamma}{3H}\right) \frac{da}{a}\right],\label{eq52}\\
T_e&=&T_0 (1+\omega_e)exp\left[-3\int_{a_0}^{a} \omega_e \left(1+\frac{\Gamma_e}{3H}\right) \frac{da}{a}\right],\label{eq53}
\end{eqnarray}

where $T_0$ is the common temperature of the two fluids in equilibrium phase and $a_0$ is the scale factor in the equilibrium state. In particular, using equations (\ref{eq46}) and(\ref{eq52}), the temperature of the normal fluid for constant $\omega$ can be written explicitly in the following form,
\begin{eqnarray}\label{eq54}
T&=&T_0 (1+\omega) \left(\frac{a}{a_0}\right)^{-3\omega} exp\left[-2 \int_{t_0}^{t}\frac{\omega (1+3\omega)}{1+\omega} \phi dt\right].
\end{eqnarray}

In general, at very early phases of the evolution of the universe  $T_e<T$ and then when the cosmic fluid attain equilibrium i.e, $a=a_0$,  one has $T=T_e=T_0$. In the next phase of evolution of universe one has $a>a_0$  and $T_e>T$ because energy flows from effective fluid to the usual fluid continuously and hence the thermodynamical equilibrium is violated. Now, from thermodynamical consideration, equilibrium temperature $T_0$ can be considered as the (modified) Hawking temperature \cite{Chakraborty:2012cw} i.e,
\begin{equation}\label{eq55}
T_0=\frac{H^2 R_h}{2\pi}\Bigg|_{a=a_0},
\end{equation}

where $R_h$ is the geometric radius of the horizon, bounding the universe.

\section{Brief discussions and concluding remarks}

A detailed cosmological study for gravity with torsion is done in present work. At first it is examined that whether a non-singular universe model is possible or not and it is found that for a particular choice of torsion field an emergent scenario is possible. Further it is shown that this gravity model is equivalent to the Einstein gravity with particle creation mechanism. Also it is possible to have a complete cosmic evolution starting from inflationary era to present late time accelerating era through the matter dominated era. Further it is shown that the modified Friedmann equations for this gravity can be considered as the Friedmann equations for the Einstein gravity with interacting two fluid system of which one is the usual fluid and other is effective fluid. The former is created and latter is annihilated in course of cosmic evolution. Also from thermodynamical consideration the temperature of individual fluid particles are evaluated and are found to be distinct. %Therefore the present modified gravity theory with torsion under consideration is equivalent to Einstein gravity with an additional fluid component interacting with the original matter component.
Lastly, different choices of hypothetical fluid component give rise to different cosmic solutions.

\section*{Acknowledgments}

The author A.B acknowledges UGC-JRF and S.C. thanks Science and Engineering Research Board (SERB), India for awarding MATRICS Research Grant support (FileNo.MTR/2017/000407).

\end{document}